\documentclass[twocolumn,showpacs,preprintnumbers,amsmath,amssymb,eps]{revtex4-1}

\usepackage{graphicx}
\usepackage{dcolumn}
\usepackage{bm}
\usepackage{mhchem}
\usepackage{graphicx}
\usepackage{setspace}
\usepackage{calc}
\usepackage[cdot,textstyle,mediumqspace]{SIunits}
\usepackage{sistyle}
\usepackage{mathptmx}
\usepackage{float}
\usepackage{amsmath}
\usepackage{color}
\usepackage{ulem}

\begin{document}


\title{Random Resistivity Network Calculations for Cuprate Superconductors
under an Electronic Phase Separation Transition\\}
\author{C. F. S. Pinheiro}%
 \email{felpin1@gmail.com}
\affiliation{Centro de Pesquisa de Tecnologia Nuclear, Belo Horizonte, MG
31270-901, Brazil}%
\author{E. V. L de Mello }
\altaffiliation[]{evandro@if.uff.br}
\affiliation{%
Instituto de F\'{\i}sica, Universidade Federal Fluminense, Niter\'oi, RJ
24210-340, Brazil\\}%

\date{\today}

\begin{abstract}

The resistivity as function of temperature of high temperature
superconductors is very unusual and despite its importance lacks 
an unified theoretical explanation. It
is linear with the temperature for overdoped compounds but it falls
more quickly as the doping level decreases. The resistivity
of underdoped cuprates increases as an insulator below a characteristic
temperature where it shows a minimum. We show that
this overall behavior can be explained by calculations using
an electronic phase segregation into two main component phases
with low and high {\it electronic }densities.
The total resistence is calculated by the various contributions
through several random picking processes of the local resistivities
and using a common statistical Random Resistor Network approach.

\end{abstract}

\maketitle
\section{Introduction}

The search for the underlying mechanism of the High critical temperature
superconductors (HTSC) has become one of
the most studied problems in condensed matter physics. Despite the
intense experimental and theoretical effort there is neither
a consensus on the basic mechanism nor on the the generic phase diagram.
In opposition to the low temperature superconductors that the normal
phase is a well known Fermi liquid, the normal phase of HTSC has
many non understood and non Fermi liquid like 
properties. For instance, they have a normal state gap
or pseudogap that in some cases remain at temperatures much higher
than the resistivity transition temperature $T_c$\cite{TS} and
whose origin is still a matter of great debate.

The pseudogap manifests itself by a suppression of spectral weight
of the normal-state electronic density of state as established by
many different experiments on bulk and surface, momentum and real
space, as has been extensively discussed in many review
articles\cite{TS,Tallon,Sawatzky}. Consequently, it has become
accepted that the HTSC phase diagram is dominated  by two energy
scale and two distinct quasiparticle dynamics\cite{Sawatzky,Tacon}.
One associated with the superconducting phase the other associate
with the pseudogap temperature $T^*$.

Another important feature of the cuprate superconductors
is the presence of inhomogeneities. A  phase separation
transition was observed experimentally a long time ago in the 
$La_2CuO_{4+y}$\cite{Jorg}. In this experiment the phase separation
of the interstitials oxygen are accompanied by an electronic phase 
separation of holes in $\ce{CuO2}$ planes. Many different
experiments have added evidences that the charge
distribution in the $\ce{CuO2}$ planes of the high temperature
superconductors
(HTSC) is microscopically inhomogeneous  like
neutron diffraction\cite{Tranquada,Bozin}, muon spin relaxation ($\mu
SR$)\cite{Uemura},
NQR and  NMR\cite{Singer,Curro}.
These experiments indicated that the inhomogeneities are
more pronounced  on the underdoped side of the phase diagram
and related it with the non Fermi liquid behavior of the normal phase.
The subject of phase separ
Recent scanning tunneling microscopy (STM) studies on Bi2212 reveal
spatial variations of the  electronic gap amplitude on a nanometer length scale
even on overdoped compounds\cite{McElroy,Gomes,Pasupathy}. The 
local density of states (LDOS) measured
in these STM experiments\cite{McElroy,Gomes,Pasupathy}
has two different forms that are possibly associated with
a segregation into
metallic and insulator nanoscopic regions\cite{Mello11}. Consequently 
the electronic phase separation  in cuprates has been the subject of
many articles\cite{Grenier,Mello03,Mello04,DDias07,Mello09,Innocenti,Kugel,Fratini}
and many books\cite{Muller,Saini,BBianconi}.

The electronic disorder of the normal phase are manifested in many
transport properties of a HTSC series, like the resistivity as function
of the temperature $R(T)$, that despite its importance,
lacks a widely accepted explanation. Here we want to show that the
particular behavior of $R(T)$ for different compounds of
a single series is a manifestation of the
intrinsic inhomogeneity of these materials. We can divide
the resistivity behavior for a system in three general types: It falls linearly
with the temperature for samples in the overdoped
regions\cite{Takagi}. It falls below the linear trend for
compounds near the optimum values as measured
by several groups\cite{Naqib,Carlos}. For very low doping
it falls down with the temperature, reaches a minimum at an
intermediate temperature, starts to increase at lower temperature 
until it reaches a local maximum, an then it drops to zero
below $T_c$.  This such a reentrant behavior is observed in
almost all underdoped cuprates\cite{Ando,Ono,Oh} but its origin has not
been very much discussed in the literature. The significance and
origin of such rich phenomena will be the focus of our work.
 
Based on the above facts we perform calculations taking
into account  the EPS transition by the use of the general phase
separation theory of Cahn-Hilliard (CH)\cite{CH}. It describes how a system
evolves from small fluctuations around an average density $p$ to a complete
separation into low and high density regions. At each of these small regions we
will associated a local resistivity derived from the work of Takagi et
al\cite{Takagi}, who performed  systematic measurements on the LSCO series
($\ce{La_{2-p}Sr_pCuO_4}$) to a large range of values of $T$ and $p$. Then we
use the Random
Resistor Network (RRN) method\cite{Kirk} to derive the system resistivity as a
function of the temperature $R(T)$ in a similar way as was
applied to the manganites\cite{Dagotto}. The local density and resistivity is
randomly
picked and the final $R(T)$ result is an average of many different
configurations.
We show that the non conventional features discussed in the former paragraph are
well described by this approach. 

\section{Electronic Phase Separation}

The electronic phase
separation was found by several  
HTSC experiments and detected to be in the form of either
stripes\cite{Tranquada,Bianconi}, patchwork\cite{Pan} or
checkerboard\cite{Hanaguri}. Earlier it was believed that the underline
mechanism of electronic phase separation  was associated with
defects, oxygen or cation disorder. Indeed, oxygen phase segregation
was observed on the $\ce{La_2CuO_{4+\delta}}$ by x-ray and transport
measurements\cite{Grenier,Jorg}. However depending on the type of
experiment, there are some HTSC materials, specially YBCO, that
appear to be more homogeneous or, at least do not display any gross
inhomogeneity\cite{Bobroff,Loram}.

On the other hand, electronic phase separation appear to be an
universal feature of HTSC. It is seen in the entire series of LSCO
by neutron diffraction\cite{Bozin} and in many BSCO samples by
STM\cite{McElroy,Gomes,Pasupathy}. ARPES have also detected distinct
quasiparticle behavior in LSCO and BSCO. Recently inhomogeneous
magnetic-field response measurements on some LSCO compounds and on
the $\ce{YBa_2Cu_3O_y}$ (YBCO) family\cite{Hardy} were also interpreted as
compatible
with electronic phase separation in form of domains. Specially in
the case of YBCO where doping occurs via a  change in the oxygen
concentration of the CuO-chain, the inhomogeneities cannot be
attributed to disorder in the cations substitution, providing strong
support to an intrinsic EPS transition. The effect of the
annealing time connected to the structural organization of 
oxygen interstitials in $La_2CuO_{4-y}$ was studied recently
X-rays diffraction techniques\cite{Fratini}

The possible origin of this EPS transition is the
proximity to the insulator AF phase, common to all
cuprates, as we derived from the principle of the competing
minimum free energy\cite{Mello09}. These calculations yield
a line $T_{PS}(p)$ close to the experimental results
reported by Timusk and Statt\cite{TS} ($T^0$ in their notation),
and it has been taken as a linear function
just for simplification as it is schematically shown in 
Fig.(\ref{PSDiag}) and (\ref{fig:meudiagrama}). 
When the 
temperature decreases below $T_{PS}(p)$, the free energy of the
homogeneous system with average density $p$ becomes lower than the anisotropic
one made of a bimodal distribution~\cite{Mello03} of AF domains
with $p(i) \approx 0$ and high hole density domains with
$p(i)\approx 2p$. This approach is justified by the NQR
experimental data\cite{Singer} which was interpreted as 
coming from two distinct
regions with symmetric local doping around the average value
$p$. The two symmetrical local dopings increase their 
differences as the temperature is lowered, that is
characteristic of second order transitions. 

To describe the formation of the local hole density domains we
define the difference between the local and the average  charge
density $u(i,T)\equiv (p(i,T)-p)$ as the EPS transition order
parameter. Clearly $u(i,T)=0$ corresponds to the homogeneous system
above $T_{PS}(p)$. Then the typical Ginzburg-Landau free energy
functional in terms of such order parameter near the transition is
given by

\begin{eqnarray}
f(i,T)= {{{1\over2}\varepsilon^2 |\nabla u(i,T)|^2 +V(u(i,T))}}.
\label{FE}
\end{eqnarray}
Where the potential ${\it V}(u,T)= A^2(T)u^2/2+B^2u^4/4+...$,
$A^2(T)=\alpha(T_{PS}(p)-T)$, $\alpha$ and $B$ are constants that
lead to lines of constant values of $A(T)/B$, parallel to
$T_{PS}(p)$, as shown Fig(.{\ref{Vu}). $\varepsilon$ gives the
size of the grain boundaries among two low and high density phases
$p_{\pm}(i)$\cite{Otton,Mello04}. The energy barrier between two
grains of distinct phases is $E_g(T)=A^4(T)/B$ that is proportional
to $(T_{PS}-T)^2$ near the transition, and becomes nearly constant
at low temperatures. The two minima of the  ${\it V}(u,T)$ is shown
in Fig.(\ref{Vu}.

\begin{figure}[!ht]
\includegraphics[height=6cm]{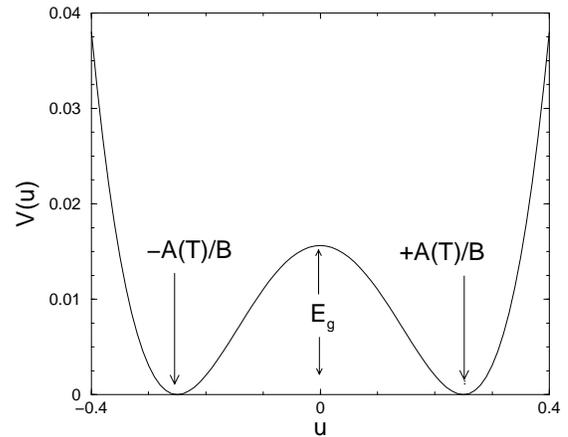}
\caption{ 
The potential commonly used in the free energy of the CH equation which gives rise to
phase separation as function of the order parameter $u$ or the density $p$. 
Notice that the two minima  yield the two equilibrium low and high density
phases and the energy barrier or free energy wall is  $E_g$ and depends on
how low the temperature is below $T_{ps}$, that is,
the temperature difference $T_{ps}-T$.}
\label{Vu}
\end{figure}

Thus, hereafter
we will use $E_g(p,T)\equiv V(p,T)$ as the grain boundary potential
or the binding potential for holes in a given grain.
$V(p,T)=V(p)\times V(T)$ and we assume, for simplicity, that $V(p)$
has a linear behavior on $p$, whose equipotentials are parallel to
$T_{PS}(p)$. In Fig.(\ref{PSDiag}) we plot $T_{PS}(p)$ where the phase
separation starts, $T_0(p)$ where the EPS is detected and $T_1(P)$ 
where insulators grains appears, 
all assumed to be linear on $p$ just for simplification but in
agreement with some signal detected by many experiments\cite{TS,Tallon,Sawatzky,Tacon}.
DM and DI are phases with low disorder in the sense that the two
phases are metallic (DM) or insulator (DI). At lower temperatures,
the disorder increases and the system is composed of very low
density or insulator regions and very high or metallic regions
(DM+DI). See Fig.(\ref{PSDiag}

The CH equation can be written\cite{Bray} in the
form of a continuity equation of the local density of free energy $f$,
$\partial_tu=-{\bf \nabla\cdot J}$, with the current ${\bf J}=M{\bf
\nabla}(\delta f/ \delta u)$, where $M$ is the mobility or the
charge transport coefficient. Therefore,
\begin{eqnarray}
\frac{\partial u}{\partial t} = -M\nabla^2(\varepsilon^2\nabla^2u
+ A^2(T)u-B^2u^3).
\label{CH}
\end{eqnarray}

In previous papers we have verified the applicability of the CH
approach\cite{Otton}. We have made a detailed study of the density profile
evolution in a $105\times 105$ array as function of the time steps,
up to the stabilization of the local densities, for parameters that
yield stripe\cite{DDias07} and patchwork\cite{Mello04,Mello08}
patterns. 

Here we want to concentrate on the energy structure of the
grains which leads to additional scattering and changes the whole
charge dynamics. We show that below the $T_{PS}$, the grain boundary
energies $E_g$ (see the inset of Fig.(\ref{PSDiag})) 
increase as the temperature goes down, segregating and confining holes at
certain regions or grains with an increasing difference in the local doping. 
The confining potential favors pair formation at
low temperatures, giving rise to the local superconductivity in
isolated regions. 
\begin{figure}[!ht]
\includegraphics[height=6cm]{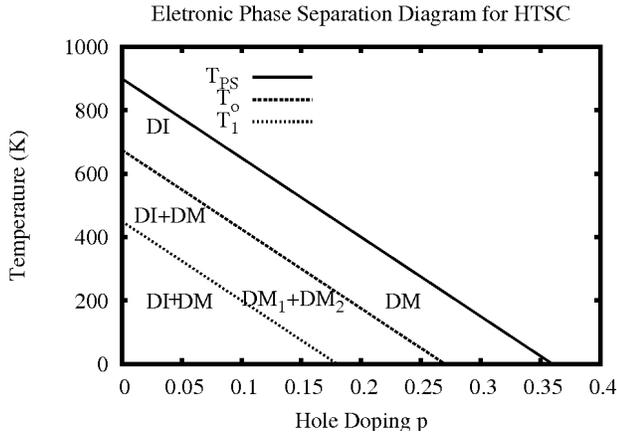}
\caption{ Important crossover lines  derived from many
experiments\cite{TS} that are associated with the degree of phase
separation. The electronic segregation starts at $T_{PS}(p)$ resulting into
two types of phases, high and low densities as a consequence of the 
potential shown in Fig.1. DI and DM 
stands for disordered insulator and disordered metal phases.
DI+DM stands for a larger disorder with insulator and metallic
phases cohexisting. $DM_1+DM_2$ is the situation which both phases 
are metallic, but with different densities.
}
\label{PSDiag}
\end{figure}

As the temperature goes down below the $T_{PS}(p)$, the system
increases the phase separation and in favorable conditions tends to
form a bimodal distribution made of low and high density domains.
Fig.(\ref{PSDiag}) shows the various levels of phase separation as
function of the temperature difference $|T_{PS}-T|$. At low density
($p \le 0.16$) and low temperature the system can achieve the
maximum charge segregation and become composed essentially of two
densities, one of $p(i) \approx 0$ and other of $p(i) \approx 2p$.
There are disordered insulator (DI) regions and disordered metallic
(DM) ones. In Fig.(\ref{Map6000}) we show this situation for the
case with $p=0.16$ and $T \approx 0$K, with half of the sites with
$p(i) \approx 0$ and the other half with $p(i) \approx 0.32$, as it
is seen in the inset histogram made over all the sites ``~$i$~" of the
system.

The density map shown in Fig.(\ref{Map6000})  
is made on a $100 \times 100$ unit cells
square lattice, the low and high density grains have different size and are
composed by 30-100 cells.  The inset is the histogram of the local doping at
each crystalline site that shows how the density
segregates into two main bands whose width values depend on
temperature.

At low temperatures below $T_0(p)$ (see Fig.(\ref{PSDiag})),
we can consider the grains as isolated entities with very weak intergrain
coupling and a typical Fermi
energy of a grain is much smaller than the bulk Debye energy, the so called
anti adiabatic regime\cite{JR97}. 

Thus below the $T_0$ line in Fig.(\ref{PSDiag}), a cuprate is
composed of a mixture of insulator ($p(i)\approx 0)$ and metallic
($p(i)\approx 2p$) grains. A clear experimental realization of such
system is the granular Pb film on a glass substrate which can change
from insulator to superconducting behavior with increasing Pb
grains\cite{Merchant}. For very low average doping  the system
is mainly insulator, but for $p\ge 0.05$ the large density grains become
metallic like, and in general both types of regions are present in one
single sample. 

\begin{figure}[!ht]
\includegraphics[height=6cm]{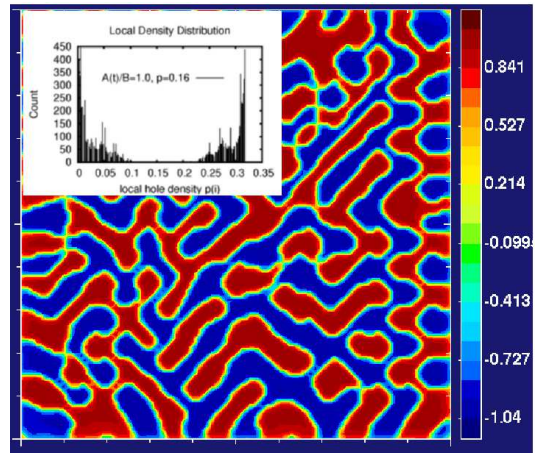}
\caption{(color online) Density map of system 
at low temperatures. The low and high density domains are clearly
formed. The inset is the local density $p(i)$
histogram with average hole density $p=0.16$, showing
that the high density grains
are metallic like and the low density (insulator) ones are
made of $p(i)\approx 0$.}
\label{Map6000}
\end{figure}

As the temperature decreases below $T_{on}(p)$ 
(see Fig.\ref{fig:meudiagrama}) it is possible to show that
the d-wave superconducting amplitude $\Delta_d(i,T)$ appears in
some isolated metallic grains\cite{Mello09}, what is known as intragrain
superconductivity by the theory of granular superconductors\cite{Merchant}.
And the zero resistivity transition takes
place only when the Josephson coupling $E_J$ among these grains is
sufficiently large to overcome thermal fluctuations\cite{Merchant}, that is,
$E_J(p,T=T_c) \approx k_BT_c(p)$ what leads to phase locking and
long range phase coherence. 

In the next section we will use the above results to calculate
the average resistivity. The main point is that any
compound of a HTSC
system goes through a EPS transitions and the local densities
follow a bimodal distribution with a dispersion similar 
as that shown in
the insert histogram of Fig.(\ref{Map6000}). This 
is interpreted as a distribution of local resistivities and will
be used in connection with the RRN\cite{Kirk}.

\section{The Resistivity Calculations}

The RRN has been quite useful in the study of systems with this
type of disorder like on the percolation phenomenon
as well as conductance models themselves\cite{AndradeJr.PRB96}. Here we adopt a
square lattice, where
nodes are connected by resistive links, which can have different resistances.
Each of these would be a nanoscopic domain inside our sample, which could be
metallic, insulating (AF phase) or superconducting. The distribution of their
fractions depend on temperature and nominal doping composition. Two bus bars
are imagined to be placed at top and bottom of the array so that a external
voltage source can set a voltage drop $V$ across the network. The total
resistivity will be simply obtained by the ratio $V/I$ where
$I$ is the total current through circuit.

As we discussed in the previous section, above the phase separation line (PS),
the
system has only metallic links, and resistance is considered to be a linear
function of temperature. Below PS, phase separation begins and two bands of
composition $p$ are allowed. These bands have their central values $\Delta p$
apart and are $\delta p(i)$ wide. Both $\Delta p(i)$ and $\delta p(i)$ depend on
temperature and are all symmetric to the nominal or average composition $p$. In
Fig.(\ref{fig:meudiagrama}) the bands are drawn in a schematic phase diagram
for a nominal doping fraction $p=0.07$. The temperature evolution of these
bands follows the study of the histograms at different temperatures as
the one shown in the inset of Fig.(\ref{Map6000}).

\begin{figure}[!ht]
 \centering
 \includegraphics[width=0.9\columnwidth,clip,angle=0]{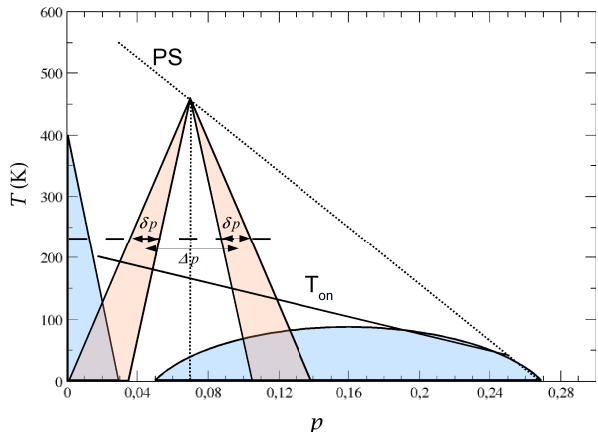}
 \caption{(color online) Schematic phase diagram showing the EPS with bands of 
doping fractions $\delta p(i)$ as a
function of temperature. 
At low temperature below $T_{on}$ tiny domains of SC appear on the right band, 
and for low doping compounds small domains of AF insulators appear on the left 
band inside the sample.}
 \label{fig:meudiagrama}
\end{figure}

Resistivity is a function of both composition and temperature. From the
results of Takagi et al\cite{Takagi}, for various compositions $p$ and a given
temperature, is possible to devise a nearly exponential
dependence on composition. The metallic links in the RRN will
then follow a random distribution in the allowed values of $p(i)$
(the bands) and their resistances will be given by a function derived
from the values of Takagi et al\cite{Takagi}
\begin{equation}
 R(p(i),T)=A(T)\exp\{-B(p(i)-p)\}
 \label{eq:RpT}
\end{equation} 
Where B(T)=0.05 and A(T) are derived directly from the LSCO
series measurements\cite{Takagi}.

On the other hand, following experimental results\cite{Pasupathy,Carlos,
Merchant} and calculations\cite{Mello09,Mello03} similar to
the intragrain superconductivity discussed in the previous
sections, some superconducting regions appear at low temperature.
Thus, as the temperature decreases below $T_{on}(p)$, 
a fraction of the most conducting metallic
links (largest $p(i)$'s) are replaced by SC links and this fraction
reaches 100\% of the
metallic band at $T_c(p)$.
Similarly, insulators appear in the low density band as $p\rightarrow
0$, which happens at lower temperatures, as can be seen in
Fig.(\ref{fig:meudiagrama}).
Likewise the CH phase separation calculations shown in the previous section,
the resistivity calculations were made on $100\times 100$ lattices. 
Periodic boundary conditions were used only on the sides without the $V$ source
terminals. The resulting linear systems
that come from Kirchhoff laws were solved with the aid of the subroutine
\texttt{ma57}, provided by HSL, that makes use of a multifrontal
algorithm\cite{MA57},
what speeded up the calculations. To each value of $p$ and $T$ we use
300 samples in order to obtain results independent of the random
process of sharing the $p(i)$ distribution.

\begin{figure}[!ht]
 \begin{center}
 \includegraphics[width=0.9\columnwidth,clip,angle=0]{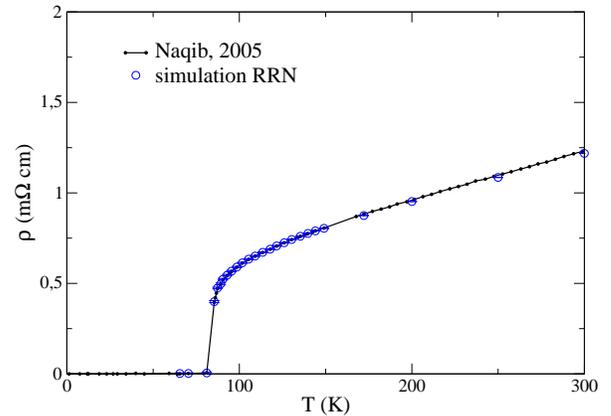}
 \end{center}
 \caption{(color online) Simulation results compared with experimental data for  the case 
$\ce{Y_{0.80}} \ce{Ca_{0.20}} \ce{Ba_2 Cu_3} \ce{O_{7{-}\delta}}$ with average
doping $p=0.136$. As the temperature decreases below $T_{on}=160K$ some of
the resistance become superconducting   and total resistivity decreases
below the linear regime. An average with 300 samples were 
used for each point in the graph.}
\label{fig:resultnaqib}
 \end{figure}
Results of simulations for typical samples near optimum 
doping are shown in Fig.\ref{fig:resultnaqib} for the compound 
$\ce{Y_{0.80} Ca_{0.20} Ba_2 Cu_3 O_{7{-}\delta}}$ with nominal doping
$p=0.136$. 300 samples were used for each
point in the graph. The phase separation begins at \SI{400}{K}; SC links of
resistivity \SI{1e-8}{m\ohm.cm} are introduced below \SI{180}{K}, 
according to the $T_{on}(p)$ line.

The small error bars can tell how reliable the results are, 
that is, a very small dispersion is attained with averages made on 300 configurations. It
is also remarkable how the $\rho\times T$ linear dependence is attained 
in the calculations with the two bands scenario.
That can be explained by the proper choice of the function
$R(p,T)$ defined in
Eq.~(\ref{eq:RpT}), which can be better understood under the light of results
obtained by Costa \textit{et al.} who used renormalization group theory to
study square RRN~\cite{Costa-1986}. When a square RRN has resistors of only two
conductivity
values $g_1$ and $g_2$, distributed with probabilities $q-1$ and $q$,
respectively, Costa \textit{et al.} showed that the network overall
conductance, expressed as a function of probability $q$, is
\begin{equation}
 \frac{\sigma(q)}{\sigma(1)} \frac{\sigma(1-q)}{\sigma(1)} = \frac{g_1}{g_2}
 \label{eq:tsallis1}
\end{equation}  
With equal probabilities, i.e. $q=0.5$, its overall resistance will simply be
\begin{equation}
 r=\sqrt{r_1r_2}.
\label{eq:tsallis2}
\end{equation} 

In our case, if the widths $\delta p$ were chosen to be zero, a network of links
with resistances $R(p-\Delta p(i),T)$ and $R(p+\Delta p(i),T)$ would result in
an overall resistance $R(T)=A(T)$, easily derived from Eqs.
(\ref{eq:RpT}) and (\ref{eq:tsallis2}). The same result holds for finite width
$\delta p(i)$, if symmetric distributions around $p$ are used. Making $A(T)$
linear is then the right choice. Departures from this linear behavior are
explained by the growth of SC and AF domains.

The exact linear dependence of $A(T)$ for a given doping fraction $p$
can be extracted from the linear part of experimental curves as in 
Figs.(\ref{fig:resultnaqib} and \ref{fig:takagi_p=07})

The fraction of SC links to be inserted below $T_{on}$ 
was also calculated from
Eq.~\ref{eq:tsallis1}, as follows. The difference between function $A(T)$ and
experimental data was normalized by $A(T)$ itself, and Eq.~\ref{eq:tsallis1}
gave the probability $q$ associated with a network of conductances $g_1=1$ and
$g_2=\SI{1e+8}{m\ohm.cm}$. When distributing $p(i)$ in the allowed bands by
means
of random numbers $x$, uniformly distributed in the interval $[0,1]$, those
links for which $x>(1-q)$ (the best conductors) were made superconductors.

The resistivity for low doping materials requires special
care because it may contain insulator, metallic and superconducting
regions represented by links in the RRN method. Thus 
simulations for a case which  reentrant behavior is present were
made for the $p=0.07$ compound and the results are shown
in Fig.~\ref{fig:takagi_p=07}. We find that
such complex behavior is due to the mixture of superconducting 
and insulator regions in the two different bands. 
The appearance
of the insulator regions in our RRN matrix occurs
when $\delta p(i)$ crosses the value of $p(i)\le0.05$\cite{Oh} at low
temperatures as it is demonstrated in Fig.~\ref{fig:takagi_p=07}.
This is the reason why,
at low temperatures, the resistivity increases,
but eventually, as  the temperature decreases further and
crosses $T_{on}$, 
the metallic resistors become superconductors and
the resistivity vanishes due to the percolation among the superconducting
regions, as we explain below.

\begin{figure}[!ht]
 \begin{center}
  \includegraphics[width=0.8\columnwidth,clip]{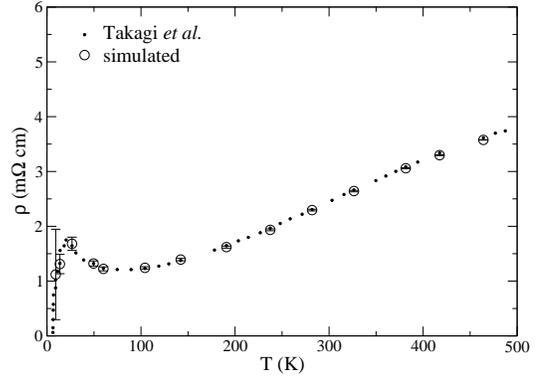}
 \end{center}
 \caption{Simulation results (circles) for in-plane resistivity of
$\ce{La_{2{-}p}Sr_pCuO4}$ with  $p=0.07$ and the reentrant
behavior. Each calculated point
is the average of 600 configurations. Dots represent 
experimental data obtained by Takagi \textit{et al.} \cite{Takagi}.
}
 \label{fig:takagi_p=07}
\end{figure}
The appearance of the superconducting links 
(resistivity \SI{1e-8}{m\ohm.cm}) begins when the temperature decreases
below 280~K. Their fraction is derived in a similar fashion, based on
 Eq.~(\ref{eq:tsallis1}), as described above. Insulating links
($\rho=\SI{200}{m\ohm.cm}$) were first added at 140~K as
$p(i)$ reaches the value 0.05 in the low density band. 
In this case, due to the low values of the densities, the number of samples for
each point was 600 for temperatures below 60~K and 300 for the remaining
temperatures. As $T \rightarrow 0$ both insulating and SC links fractions come
close to the link percolation threshold 0.50. That is the reason for the strong
variation
expressed in the error bars as $T \rightarrow 0$ in Fig.~\ref{fig:takagi_p=07}.

\section{Conclusion}

In this paper we derived the resistivity of a complete HTSC series 
assuming an EPS transition. 
Following many experimental results we took into consideration that a HTSC 
compound undergo an electronic segregation and is composed mainly of two types
of regions or grains. These are the high and low 
local density which gives a metallic or insulator behavior
to the grains as observed by the STM measurements of the LDOS.
We used simulations with the CH theory to follow the effects
of the EPS as function of the temperature in order to describe
the disordered composition of a HTSC sample.

 With this approach the overall
resistivity can be calculated through the RRN method\cite{Kirk}
and used to reproduce the measured values either in the 
overdoped or underdoped regions. Our method based on the
effects of disorder is in very good agreement with the measured
departure of the linear behavior of near optimum compounds and
the completely different down and up reentrant behavior for weakly doped
samples that have been studied in detail\cite{Oh}, but so far without
a simple and unified interpretation. 

The excellent fitting on 
such a rich variety of features with one single approach 
without adjustable parameters and with
values of the local resistivities taken from the measurements
of Takagi et al\cite{Takagi} for an entire
series, let us conclude the following:  
the EPS described here  and the onset of local orx
intragrain superconductivity given by the curve $T_{on}$ 
are  universal 
properties and must be considered
to interpret the transport properties of any HTSC system.

\section{Acknowledgment}
We gratefully acknowledge partial financial aid from Brazilian
agencies CNPq and Capes.

\end{document}